\documentclass[aps,prd,twocolumn,showpacs,preprintnumbers,amsmath,amssymb,nofootinbib]{revtex4-1}
\usepackage{amsmath,amsfonts,euscript,amssymb}
\usepackage[cp1251]{inputenc}
\usepackage{colordvi}
\usepackage{color}
\usepackage[T1]{fontenc}
\usepackage[english,russian]{babel}
\usepackage{graphicx}
\usepackage{bm}
\newcommand{\be}{\begin{equation}}
\newcommand{\ee}{\end{equation}}
\newcommand{\bea}{\begin{eqnarray}}
\newcommand{\eea}{\end{eqnarray}}

\begin{document}

\title{Von Neumann's Quantization of General Relativity
}

\author{A.B. Arbuzov$^{1,2,\diamondsuit}$,
A.Yu. Cherny$^{1}$,
D.J. Cirilo-Lombardo$^{1,3}$
R.G. Nazmitdinov$^{1,4}$,
Nguyen Suan Han$^{1,5}y$
A. E. Pavlov$^{1,6}$,
V.N. Pervushin$^{1}$,
A.F. Zakharov$^{1,7}$
}
\affiliation{${}^1$Bogoliubov Laboratory of Theoretical Physics,
Joint Institute for Nuclear Research,
Joliot-Curie str. 6, Dubna, 141980, Russia \\
$^2$Department of Higher Mathematics, Dubna State University,
Dubna, 141980, Russia \\
$^3$National Institute of Plasma Physics, (INFIP-CONICET), FCEyN,
Universidad de Buenos Aires, Buenos Aires, 1428, Argentina \\
$^{4}$Department de F{\'\i}sica, Universitat de les Illes Balears,
Palma de Mallorca, E-07122, Spain \\
${}^{5}$ Hanoi University of Science,Vietnam National University, Hanoi, Vietnam \\
$^{6}$Institute of Mechanics and Energetics, Russian State Agrarian University,
Timiryazevskaya, 49, Moscow 127550, Russia\\
${}^{7}$National Research Nuclear University  MEPhI, 
115409, Moscow, Russia\\
$^{\diamondsuit}$arbuzov@theor.jinr.ru
}

\begin{abstract}
Von Neumann's procedure is applied for quantization of General Relativity.
We quantize the initial data of dynamical variables at the Planck epoch,
where the Hubble parameter coincides with the Planck mass.
These initial data are defined via the Fock simplex in the tangent Minkowskian space-time
and the Dirac conformal interval. The Einstein cosmological principle
is applied for the average of the spatial metric determinant logarithm over
the spatial volume of the visible Universe. We derive
the splitting of the general coordinate transformations into
the diffeomorphisms (as the object of the second N\"other theorem)
and the initial data transformations (as objects of the first N\"other theorem).
Following von Neumann, we suppose that the vacuum state is a quantum ensemble.
The vacuum state is degenerated with respect to quantum numbers of non-vacuum states
with the distribution function that yields the Casimir effect in gravidynamics
in analogy to the one in electrodynamics.
The generation functional of the perturbation theory in gravidynamics
is given as a solution of the quantum energy constraint.
We discuss the region of applicability of gravidynamics and its possible predictions
for explanation of the modern observational and experimental data.
\end{abstract}

\date{}



\maketitle

\section{Introduction
\label{sect_1}}

In 1915 Albert Einstein introduced the field equations of
gravity~\cite{Einstein:1915ce}.
The mathematical basis of these equations is their invariance
with respect to the general coordinate transformations in the Riemannian space-time.
Solutions of these equations express the
Riemannian space metric components through the matter energy-momentum ones.
These solutions contain additional constants known as certain initial data.
Einstein supposed that the general coordinate transformations have the same physical
status as the initial data transformations under the Galileo group or its
relativistic generalization known as the Poincar\'e symmetry group
in the Special Relativity (SR).

In the same 1915 David Hilbert derived the Einstein equations by variation
of the corresponding action with respect to the metric components~\cite{Hilbert}.
The Hilbert action contained new information in comparison with the equations.
Along this line,  Emmy N\"other proved two theorems~\cite{Noether-18}.
In accordance with these two N\"other's theorems, there are two different symmetry groups.
The first of them is the global Poincar\'e  group of the initial data transformations as the object
of the first N\"other theorem about the conservation laws that follow from a global group.
The second symmetry group is the set of general coordinate transformations
({\it i.e.} diffeomorphisms) as the object of the second N\"other's theorem asserting
that the time-time and time-space components of the Einstein equations are constraints 
of the initial data. Therefore one should distinguish between the Einstein diffeomorphisms 
and the initial data transformations.


The difference between the Einstein diffeomorphisms and
the transformations of reference frames was found in the approach to the
General Relativity (GR) developed by Vladimir Fock~\cite{Fock-1929}.
Fock introduced a diffeo-invariant orthogonal basis in order
to include fermions in the GR. In this way, he replaced the metric
components in the GR by the diffeo-invariant simplex components in the
tangent Minkowskian space-time. By using the Fock's modification of
the Einstein's approach, one can introduce the Poincar\'e symmetry group 
of the initial data transformations in the tangent space-time.

In the framework of the Hilbert's ``Foundations of physics'' \cite{Hilbert},
the next steps forward in the conception of the initial data transformations were
made by Dirac~\cite{Dirac-1958} and by Arnowitt, Deser, and Misner~\cite{ADM-1959}.
Namely, they formulated the so-called Dirac--ADM Hamiltonian approach to constrained
gravitational systems described by the Hilbert action. In this case, the Einstein
diffeomorphism group transformations are reduced to
its kinemetric subgroup ones~\cite{zelmanov-1989} that include
the Riemannian time reparameterizations.
The reparameterization time invariance 
means that there are two invariant evolution parameters:
the \textit{geometrical} and \textit{dynamical} ones.
The first parameter is the geometrical interval, and
the second can be associated with the zeroth mode of one of
the dynamical fields in the Wheeler--DeWitt (WDW)
field superspace of events~\cite{DeWitt,Wheeler}.
Then the \textit{energy of events} in the WDW field superspace space
is a solution of the energy constraint with respect to the canonical momentum of
the \textit{dynamical} time-like field.

Ch.W.~Misner~\cite{Misner-59} supposed that the dynamical evolution time
can be identified with the cosmological scale factor logarithm that
can be considered as the zeroth mode of the spatial metric determinant
logarithm in the Dirac--ADM Hamiltonian approach to the constrained gravitation
system. The scalar field zeroth mode is defined by means of
averaging over finite spatial volume in accordance with the
cosmological principle suggested by Einstein in 1917~\cite{Einstein:1917ce}. 
Misner's definition of the dynamical evolution time by means of
the cosmological principle becomes diffeo-invariant, if we restrict spatial diffeomorphisms
by uniform transformations that preserve the spatial metric determinant.

Following Dirac~\cite{Dirac:1973gk}, one can call the Einstein spatial
metric determinant logarithm as a \textit{dilaton} field. Dirac and
Deser~\cite{Deser-70} replaced the Einstein intervals by the dilaton-free 
conformal intervals and identified these intervals  with the
observable distances. In this case, the dilaton field describes
both the redshift in Cosmology and relativistic effects, such as the
Mercury anomaly or double Eddington angle of the photon deviation
in the Sun gravitational field.
If the Hubble parameter and all masses are equal zero,
then the Einstein's theory becomes conformal invariant.
Therefore, one can say that the dilaton field
is the Goldstone mode accompanying the spontaneous conformal symmetry breaking.

The Goldstone mode approach to the Einstein GR was developed in paper~\cite{Borisov:1974bn}.
The authors obtained the Hilbert action as a joint nonlinear realization of
the affine symmetry group of all linear transformations [A(4)] and conformal
symmetry transformations [C=SO(4,2)] by analogy with the Schwinger--Weinberg
phenomenological Chiral Lagrangian as a nonlinear realization of the chiral symmetry
in hadron physics~\cite{Volkov:1969bn}.
This nonlinear realization supposed that the dilaton
and all metric components are Goldstone modes. This approach collects 
all previous modification discussed before
(including  Hilbert's action, Fock's simplex, and Dirac's dilaton)
and distinguishes the so called normal coordinates in the field
of the Goldstone mode space along geodesic lines.

Papers~\cite{Barbashov:2005hu,Arbuzov:2010fz,Pervushin:2011gz}
adduce theoretical and observational arguments in favor of the separation
between the initial data symmetry group and the subgroup of the general
uniform coordinate transformations (D-diffeomorphisms).
In the present paper we use this separation in order to construct the perturbation
theory generating functional in terms of the D-diffeo-invariant independent variables
and observable coordinates.

The features of our approach to the GR are:

1) the choice of normal coordinates along geodesic lines
in the field space of the Goldstone modes,
({\it i.e.} the exponent parameterization of the diagonal metric
component~\cite{Volkov:1969bn});

2) the splitting of the Einstein (E) general coordinate transformations
onto the Dirac (D) diffeomorphisms and the initial data transformations;

3) introduction of the D-diffeo-invariant independent variables and coordinates.

These features allow us to construct unambiguously the diffeo-invariant Wheeler-DeWitt
equation and solve it
by the analogy with the irreducible unitary representations of
the Poincar\'e group~\cite{wigner-39}.

The Dirac definition of observable quantities (as the D-diffeo-invariant ones) distinguishes
between physical approximations in terms of invariants and
a mathematical ansatz beyond the invariants.
There are two approximations: \textit{static} and \textit{dynamical}.
In the \textit{static} one we neglect
the independent variables in order to obtain the static interactions of external sources.
The basic order of the \textit{dynamical} approximation means
that far from  external sources we can keep only D-diffeo-invariant independent variables.
The next orders of the \textit{dynamical} approximation
represent the \textit{gravidynamics} perturbation theory.

In the following we consider both these cases and compare the obtained results
with the classical homogeneous Cosmology as a Friedmann model of the Universe
that takes into account only the dilaton zeroth mode associated with the cosmological scale factor.
The structure of the paper is as follows. Section~II defines \textit{gravidynamics} as
the Einstein General Relativity  in terms of D-diffeo-invariant coordinates
and independent variables with corresponding initial data.
Section~III is devoted to the generating functional of the quantum  perturbation theory
as quantum solutions of the energy constraint in the GR.
Section~IV  defines the basic approximation of the quantum \textit{gravidynamics} (QGD).
Section~V estimates the predictable possibilities of the QGD
in the context of last observational facts and phenomena.

%

\section{Diffeo-invariant initial data in General Relativity}

\subsection{Nonlinear realization of the initial data symmetry group}

The history of equations of motion is associated with the names
of Newton, Maxwell, Einstein, Klein-Gordon, Dirac, and the authors of the Standard Model
and the modern string theory. However, physical facts and phenomena are
described by solutions of those equations that contain initial data.
The initial data history is more simple than the history of equations.
Initial data symmetries can be associated with only the 10-parametric
Galileo group transformations and the Poincar\'e ones obtained from Galileo ones
via the replacement of velocity translations by the Lorentz antisymmetric rotations
\bea\nonumber
\underbrace{{v_i\to v_i+v_i^0}}_{\rm {Galileo}}
\to \underbrace{{x_i\to x_i+ L_{i0}t,\,
t\to t+L_{0i}x_i}}_{\rm {Lorentz~boost}}.
\eea
In any case all the above mentioned equations for free fields can be classified
in the form of the unitary irreducible representations of the
Poincar\'e group obtained by Wigner~\cite{wigner-39}.
In other words, Wigner postulated the priority of the initial data transformation
groups and derived all equations as invariant structure relations
of the initial data transformation groups.

The Wigner construction can be extended onto the 20-parametric affine group A(4).
The affine group A(4) generalizes the 10-parametric Poincar\'e group
(that contains 4 translation generators $\hat P_{(\alpha)}$ and
6 Lorentz ones $\hat L_{(\alpha)(\beta)}$)
by 10 proper affine generators $\hat R_{(\alpha)(\beta)}$.
The latter yield 10 symmetric linear transformations of the Minkowskian space-time
coordinates
$${x}^\mu\to \widetilde{x}^\mu=
\underbrace{{{x}^\mu+y^\mu+L_{[\mu\nu]}{x}^\nu}}_{\rm {Poincare}~[10]}
+\underbrace{{R_{\{\mu\nu\}}{x}^\nu}}_{[10]}.$$
Ogievetsky proved that the 20-parametric affine group A(4) and
15-parametric conformal one are a closer of the general coordinate
transformation group \cite{Ogievetsky-73}. Ogievetsky and Borisov identified 4-coordinates
$x^\mu$ and 10 proper affine parameters $h_{\mu\nu}$ with the Goldstone modes.
They considered the motion of the orthogonal simplex in the coset $K=\dfrac{A(4)}{L}$,
where (L) is the Lorentz subgroup \cite{Borisov:1974bn}.
The shifts and the rotation of the simplex are given by  the Maurer-Cartan linear differential
forms $\omega^P_{(\alpha)}$,  $\omega^{R}_{(\alpha)(\beta)}$, and $\omega^L_{(\alpha)(\beta)}$, respectively.
The dependence of these Maurer-Cartan forms on the Goldstone modes $h^{\mu\nu}$ is determined by the
affine group algebra of commutation relations using the transformations
\bea \nonumber
&& G=e^{\imath\hat P\cdot x}e^{\imath\hat R\cdot h},
\\ \nonumber
&& G^{-1}d G \!=
\\ \nonumber
&& \imath\left[\!\hat P_{(\alpha)}\omega^P_{(\alpha)}\!+\! \hat R_{(\alpha)(\beta)}\omega^R_{(\alpha)(\beta)}
 \!+\!\hat L_{(\alpha)(\beta)}\omega^L_{(\alpha)(\beta)}\!\right],
\\ \nonumber
&& \omega^P_{(\alpha)}[h]= {\bf e}_{(\alpha)\mu}dx^\mu,
\\ \nonumber
&& \omega^L_{(\alpha)(\beta)}[h]=(1/2) \{{\bf e}_{(\alpha)\mu}d{\bf e}^\mu_{(\beta)}-
{\bf e}_{(\beta)\nu}d{\bf e}^\nu_{(\alpha)}\},
\\ \nonumber
&& \omega^R_{(\alpha)(\beta)}[h]=(1/2) \{{\bf e}_{(\alpha)\mu}d{\bf e}^\mu_{(\beta)}+
{\bf e}_{(\beta)\nu}d{\bf e}^\nu_{(\alpha)}\}.
\eea
These linear forms and the conformal symmetry principle yield unambiguously
the Hilbert action of Einstein's gravitational
theory\footnote{Here and in the following we use the natural units
$ M_{\rm Pl}\sqrt{3/(8\pi)}\equiv M^*_{\rm Pl}=c=\hbar=1$.}
\be \label{H-1a}
W_{\rm H}=-\int d^4x\sqrt{-g}\dfrac{R^{(4)}(g)}{6}
\ee
with the interval expressed in terms of the Fock simplex components
$$ ds^2=\underbrace{g_{\mu\nu}dx^\mu dx^\nu}_{Einstein (1915)}=
\underbrace{{ e^{-2D}}}_{Dirac (1973)}\underbrace{\omega_{(\alpha)}\otimes\omega_{(\beta)}}_{\rm Fock (1929)}
\underbrace{{\eta^{(\alpha)(\beta)}}}_{\rm \bf  tangent}.$$
Here we use the definition of the conformal Fock simplex components
\be\label{omega-1}
\omega_{(\alpha)}=e^D\omega^P_{(\alpha)}(h).
\ee
This nonlinear realization of affine and conformal symmetry groups not only yields the Hilbert action
but justifies of the identification of the Dirac conformal intervals
\be \label{H-2a}
\underbrace{\widetilde{ds}^2}_{\rm Dirac }=
\omega_{(\alpha)}\otimes\omega_{(\beta)}\eta^{(\alpha)(\beta)}=e^{2D}\underbrace{ds^2}_{\rm Einstein}
\ee
with the observable ones. This nonlinear realization yields
physical variables as their normal coordinates along geodesic lines in the coset $K=A(4)/L$
in terms of  diffeo-invariants in the tangent space-time marked by the round bracket indexes
$(\alpha)$~\cite{Volkov:1969bn}.

\subsection{Diffeo-invariant 3+1 foliation}

The initial data problem supposes a concrete frame of reference known as
the Dirac--ADM foliation~\cite{Dirac-1958,ADM-1959}, $4=3+1$.
In the frame one can convince that the conformal Fock simplex (\ref{omega-1}) takes the form
\bea\label{1-4}
{{\omega}}_{(0)}=e^{-2{D}}N dx^0,
\\ \label{1-4ab}
{{\omega}}_{(b)}={\bf e}_{(b)i}[dx^i+{N}^{i}dx^0],
\eea
where $N$ is the lapse function,
\be\label{N-1}
{N}^j=N^{j\bot}+N^{j||};~~\partial_l N^{l\bot} =0
\ee
are three shift vectors, $D$ is the dilaton field, and ${\bf e}_{(b)i}$ are five
symmetric triads with the unit determinant $|{\bf e}|=1$.


The Fock simplex separates Einstein's diffeomorphisms (i.e. general coordinate transformations)
$$x^\mu \to \widetilde{x}^\mu =\widetilde{x}^\mu({x}^\mu),\,g_{\mu\nu}\to
\widetilde{g}_{\mu\nu}(\widetilde{x})=g_{\alpha\beta}\frac{d{x}^\mu}{d\widetilde{x}^\alpha}
\frac{d{x}^\nu}{d\widetilde{x}^\beta}$$
onto the Riemannian time reparameterization and uniform spatial coordinate transformations
(i.e. D-diffeomorphisms)
$$x^0 \to \widetilde{x}^0 =\widetilde{x}^0({x}^0),\,{\bf e}_{(b)j}\to \widetilde{\bf e}_{(b)j}
(\widetilde{x})={\bf e}_{(b)k}({x})\frac{dx^k}{ d\widetilde{x}^j}$$
and the initial data transformations.
The latter include shifts of the dilaton $D(x)\to D(x)+ {\rm Constant}$
and the Lorentz rotations of Fock's simplex components.

The Riemannian time reparameterization separates the time-like
dynamical evolution parameter in the field space of event.
This evolution parameter can be identified with
the zeroth dilaton mode via the volume average in terms of D-diffeo-invariant forms
\be \label{dil-1}
{\langle D\rangle} \equiv V^{-1}_0\int_{V_0} \omega_{(1)}\wedge\omega_{(2)}\wedge\omega_{(3)}  D.
\ee
according to Einstein's cosmological principle \cite{Einstein:1917ce},  here
$V_0=\int_{V_0} \omega_{(1)}\wedge\omega_{(2)}\wedge\omega_{(3)}$
is the finite D-diffeo-invariant volume.
This zeroth dilaton mode coincides with the cosmological scale factor logarithm
\be\label{dil-2}
{\langle D\rangle}=-\ln a=\ln(1+z).
\ee
This quantity is known in the observational astrophysics as luminosity.
The spatial average (\ref{dil-1}) can be called as the \textit{global} projection operator,
while the orthogonal projection operator
\bea \label{ds-2}
\overline{D}&=D&-\langle D\rangle
\eea
is a \textit{local} one. The \textit{local} functions obey the constraint
$\langle \overline{D}\rangle\equiv0$.
This classification of functions is the consequence of the D-diffeomorphisms.

\subsection{The physical content of the Hilbert action}

The direct substitution of the Fock simplex components (\ref{1-4}) and (\ref{1-4ab})
into the Hilbert action determines its diffeo-invariant physical content
according to the Dirac--ADM $3+1$ foliation~\cite{Barbashov:2005hu}:
\bea \label{1-2}
&& W_{\rm H}\ =\ \underbrace{{ W_{\rm  cosmology}}}_{{ \rm =0~for~V_0= \infty}} \
+ \ W_{\rm wave} \ + \ W_{\rm gravity},
\\ \label{1-3a}
&&  W_{\rm cosmology.}=-\int d^4x \left(\frac{d\, \langle D \rangle}{dx^0}\right)^2\frac{1}{N},
\\ \label{1-4a}
&& W_{\rm wave}\!=\!\int\limits_{}^{} \!d^4x\!\frac{N}{6}\!\!\left[\!v_{(a)(b)}v_{(a)(b)}
- R^{(3)}({{\bf e}})e^{-4D}\!\!\right]\!\!,
\\ \label{1-4abc}
&& W_{\rm gravity}\!=\!\!\!\int\limits_{}^{}\! d^4x\left[\!-\!v_{\overline{D}}^2\!
-\!\frac{4}{3}{N}e^{-7D/2}\triangle^{(3)} e^{-D/2} \!\! \right]\!\!.
\eea
Here $R^{(3)}({{\bf e}})$ is the three dimensional curvature
\bea\nonumber
R^{(3)}({\bf e}) &=& - 2\partial^{\phantom{f}}_{i} [{\bf e}_{(b)}^{i}\sigma_{{(c)|(b)(c)}}]
- \sigma_{(c)|(b)(c)}\sigma_{(a)|(b)(a)}
\\ \label{1-17}
&& + \sigma_{(c)|(d)(f)}^{\phantom{(f)}} \sigma^{\phantom{(f)}}_{(f)|(d)(c)},
\eea
where
$$
\sigma_{{(c)}|(a)(b)}\!\! = \!\omega^L_{(a)(b)}(\partial_{(c)})\!+\!
\omega^R_{(a)(c)}(\partial_{(b)})\! - \! \omega^R_{(b)(c)}(\partial_{(a)}),
$$
and
\bea\label{1-4v}
&& v_{(a)(b)}
\\ \nonumber
&& = \frac{1}{N}\left[\omega^R_{(a)(b)}(\partial_0-\partial_l N^l)+\partial_{(a)}N^{\bot}_{(b)}
+\partial_{(b)}N^{\bot}_{(a)} \right],
\\ \label{1-4d}
&& v_{\overline{D}}= \frac{1}{N}\left[\partial_0(e^{-3D})+\partial_l(N^l e^{-3D})\right]
\eea
are velocities of the gravitational waves and non-zeroth dilaton modes.
One can see that the Hilbert action describes three classes of phenomena: Cosmology, waves
and potentials of gravity.
The Hilbert action (\ref{1-2}) shows us also that its kinetic part does not depend
on the antisymmetric forms $\omega^L_{(a)(b)}$. These antisymmetric forms are not
dynamical variables and they can have zero initial data
\bea \label{1-5ab}
\omega^L_{(a)(b)}(d) &=& 0.
\eea
Following the Dirac's approach~\cite{Dirac-1958} to constrained systems,
one can impose the second class constraints
\bea \label{1-6ab}
\partial_i{\bf e}^i_{(a)}&=&\omega^R_{(a)(b)}(\partial_{(b)})=0,
\\ \label{1-6abc}
v_{\overline{D}} &=& 0.
\eea
They mean the zero initial data for the longitudinal components
and the minimal hyper-surface condition~(\ref{1-6abc}).
The Hilbert action (\ref{1-2}) distinguishes three spaces:
the Riemannian one, the tangent one, and the field one.

\subsection{Diffeo-invariant coordinates}

The cosmological part of the Hilbert action (\ref{1-2}) determines the global lapse function
\be\label{1-14}
N_0^{-1}\equiv\langle N^{-1}\rangle
\ee
as the average over a finite volume.
The Hilbert action (\ref{1-2}) expressed in terms of the linear Maurer-Cartan forms
gives a possibility to introduce the D-diffeo-invariant spatial-time coordinates
\bea \label{1-14t}
d\tau &=& N_0(x^0)dx^0,
\\ \label{1-14x}
X_{(a)}&=&x^i {\bf e}_{i(a)}.
\eea
The differential of the  diffeo-invariant spatial coordinates (\ref{1-14x})
\bea \label{1-15}
d X_{(a)}= \underbrace{{\bf e}_{i(a)}dx^i}_{\widetilde{\omega}_{(a)}}
+ x^i\underbrace{{\bf e}_{i(b)}{\bf e}^j_{(b)}}_{\delta^i_j}d{\bf e}_{j(a)}
\eea
expresses them via the spatial components of the Fock simplex in Eq.~(\ref{1-15})
and the diffeo-invariant graviton ${\omega}^R_{(b)(a)}$,
\bea \label{1-16}
\widetilde{\omega}_{(a)}\equiv{\bf e}_{i(a)}dx^i=d X_{(a)}-X_{(b)}{\omega}^R_{(b)(a)}(d).
\eea
According to the Dirac's approach to constrained systems
the equalities (\ref{1-14t}) and (\ref{1-14x}) are \textit{weak} ones.
These equalities can be used only after the variation of the Hilbert action
in terms of the Riemannian space-time coordinates.
Meanwhile the diffeo-invariant tangent space-time coordinates are needed to compare
the solutions of the equations with diffeo-invariant observational data.
According to the Dirac concept of observable quantities the observable coordinates
should be D-diffeo-invariant.
Thus, the ``Foundations of the GR'' are the Hilbert action (\ref{1-2})
and the Dirac conformal interval
\bea \nonumber
&& \widetilde{ds}^2\! =\! e^{-4D}{\cal N}^2 d\tau^2\!-\!
\left[\!dX_{(a)}\!-\!X_{(b)}\omega^R_{(b)(a)}\!+\!{\cal N}_{(a)}d\tau\!\right]^2
\\ \label{ds-1}
&=&e^{-4\overline{D}}{\cal N}^2 d\eta^2\!-\!
\left[\!dX_{(a)}\!-\!X_{(b)}\omega^R_{(b)(a)}\!\!+\!{\cal N}_{(a)}d\tau\!\right]^2.
\eea
Just these ``Foundations'' lead us to diffeo-invariant variables.
The Dirac concept of observability has very important consequences.
The diffeo-invariant coordinates (\ref{1-14t}) and (\ref{1-14x})
allow us to treat the Maurer-Cartan forms (\ref{1-16}) $\widetilde{\omega}_{(a)}$ and
$\omega_{(a)(b)}^R(\partial_{(c)})$ as D-diffeo-invariant independent degrees of freedom.
The classical initial data admit zero values of the diffeo-invariant quantities
\be \label{z-0}
\ln{\cal N}=0, \quad {\cal N}_{(a)}=0,\quad
D=0,\quad \omega_{(a)(b)}^R(\partial_{(c)})=0.
\ee
These values yield the flat metrics
\bea \label{z-0f}
\widetilde{ds}^2 = d\eta^2 - \left[ dX_{(a)} \right]^2
\eea
in accord with the correspondence principle.

However, in quantum theory this solution is not stable since
\be \label{z-0a}
\langle D\rangle\neq0,\qquad \omega_{(a)(b)}^R(\partial_{(c)})\neq 0.
\ee
Below we discuss a possibility to construct a stable solution
with the help of the uncertainty principle for the strong gravitation waves in the approximation
\bea \label{1z-0f}
\widetilde{ds}^2 = d\eta^2 - \left[ dX_{(a)}-\omega_{(a)(b)}^R(\partial_{(c)}) \right]^2.
\eea
In this case, the diffeo-invariant coordinates $X_{(a)}$ given by Eq.(\ref{1-16})
\bea \label{1-16a}
d X_{(a)}=\widetilde{\omega}_{(a)}(d)+X_{(b)}{\omega}^R_{(b)(a)}(d).
\eea
are the functionals of the wave solutions
expressed via the spin-connection coefficients $\omega_{(a)(b)}^R(\partial_{(c)})$ 
and the Fock simplex components.
The latter  satisfy the independence constraints
\be\label{1-16aa}
\widetilde{\omega}_{(a)}(\partial_{(c)})\equiv\frac{{\bf e}_{i(a)}dx^i}{{\bf e}_{j(c)}dx^j}=\delta_{(a)(c)}
\ee
of the type of the flat space ones $dx^i/dx^j=\delta_j^i$. 
Using constraints~(\ref{1-16aa}) one can rewrite Eq.~(\ref{1-16a}) in the form
\bea \label{1-16aaa}
\frac{d X_{(a)}}{\widetilde{\omega}_{(b)}(d)}=\delta_{(a)(b)}+X_{(c)}{\omega}^R_{(c)(a)}(\partial_{(b)}).
\eea
In the case of $(a)=(b)$ this form reduces to
\bea \label{1-16abb}
\frac{d X_{(a)}}{\widetilde{\omega}_{(a)}(d)}=1,
\eea
while in the case of $(a)\neq(b)$  Eq.(\ref{1-16aaa}) yields the equation
\bea \label{1-16aabb}
\frac{d X_{(a)}}{\widetilde{\omega}_{(b)}(d)}=X_{(c)}{\omega}^R_{(c)(a)}(\partial_{(b)})
\eea
that allows us to express diffeo-invariant coordinates via the gravitation waves 
as we will show below in the Section~IV.

\subsection{Diffeo-invariant variables in tangent space-time}

Both action~(\ref{1-2}) and interval~(\ref{ds-1}) are bilinear quantities
with respect to the Maurer-Cartan forms.
The proposed choice of coordinates strongly simplifies resolution of
the Einstein equations in terms of D-diffeo-invariant {\em independent} variables.
These variables are only two gravitons $\omega_{(a)(b)}^R(\partial_{(c)})$
and the dilaton zeroth mode $\langle D\rangle$.
The \textit{independence} means that they can have non-zeroth initial data after their quantization.
The quantization requires canonical momenta of these independent variables that
are calculated from the Lagrangian representation of the action $W=\int d^4x {\cal L}=\int dx^0 L$
by the standard way
\bea \label{cm-D}
P_{\langle D\rangle}&=&\frac{\partial L}{\partial (\partial_0\langle D\rangle)}
= - 2V_0\frac{d \langle D\rangle}{d\tau},
\\ \label{cm-e}
p^R_{(a)(b)}&=&{\bf e}_{(a)j}\frac{\partial {\cal L}}{\partial (\partial_0{\bf e}_{(b)j})}
= \frac{v_{(a)(b)}}{3}.
\eea
The Poisson brackets take the form
\bea \label{cm-Dp}
&& \{P_{\langle D\rangle},\langle D\rangle\}=1, 
\\ \label{cm-ep}
&& \{p^R_{(a)(b)}(x^0,{\bf x}),\omega^R_{(c)(d)}(x^0,{\bf y})(\partial_k)\}
\\\nonumber
&& = \Pi_{(a)(b)(c)(d)}~\partial_k\delta^{3}({\bf x}-{\bf y}),
\eea
where $\Pi_{(a)(b)(c)(d)}$ is the projection operator. The rest fields $\overline{D}$, and
\bea \label{ds-2n}
{\cal N}&=&\frac{N}{N_0(x^0)},
\\ \label{ds-2nl}
{\cal N}_{(a)}&=& \frac{{\bf e}_{(a)l}N^l}{N_0(x^0)}
\eea
are the D-diffeo-invariant \textit{potentials}.
The variation of the Hilbert action with respect to
the potentials yields the constraints~\cite{Dirac-1958}.
In particular, the shift vector $N^l$ leads to three momentum constraints,
the lapse function logarithm $\ln N$ yields the energy constraint.
In perturbation theory these \textit{potentials} are beyond the basic approximation,
they determine only static interactions of external sources.

\subsection{The momentum constraint}

For the explicit solution of the momentum constraint
\be \label{CC-3t}
\frac{\delta W_{\rm H}}{\delta N^l}=0
\ee
it is convenient to use the expansion
\bea \label{dv-7}
N_{(b)}&=&N^{|\!|}_{(b)}+N^{\bot}_{(b)},
\\  \label{dv-8}
\partial_{(b)}N^{|\!|}_{(b)}&=& \partial_jN^j,
\\  \label{dv-9}
\partial_{(b)}N^{\bot}_{(b)}&=&0,
\\ \label{c-6}
p_{(b)(a)} &=& p^{R}_{(b)(a)}+\partial_{(a)}
f^{\bot}_{(b)}+\partial_{(b)} f^{\bot}_{(a)},
\\
\partial_{(b)}p^{R}_{(b)(a)}&=&0,
\eea
where $f^{\bot}_{(a)}$ satisfies the equation
\be \label{c-8}
\left[\triangle f^{\bot}_{(a)}+\partial_{(a)}\partial_{(b)} f^{\bot}_{(a)}\right]
=p^{R}_{(b)(c)}\partial_{(a)} \omega^{R}_{(b)(c)}  
\ee
which follows from the momentum constraint (\ref{CC-3t})
after the substitution~(\ref{c-6}).

\subsection{The energy constraint}

The invariance of the Hilbert action (\ref{1-2}) with respect to the time-reparameterizations
means that its variation under the lapse function  $\dfrac{\delta W_{\rm H}}{\delta \ln N}=0$
is the  energy density constraint:
\be \label{1-7ab}
\dfrac{1}{{\cal N}}\left[\dfrac{d\langle D\rangle}{d\tau}\right]^2={\cal N}{\cal H},
\ee
where
\bea\label{ec-2}
{\cal H}&=&-\frac{\delta [W_{\rm wave}  +  W_{\rm gravity}]}{\delta N}\\\nonumber
&=& 6 \left[p^R_{(a)(b)}+\partial_{(a)} f^{\bot}_{(b)}+\partial_{(b)} f^{\bot}_{(a)}\right]^2\\\nonumber
&+& \frac{e^{-4D}}{6}\left[ \omega^R_{(a)(c)}(\partial_{(b)})\!-\!
\omega^R_{(b)(c)}(\partial_{(a)})\right]^2
\\\nonumber
&-&\frac{4}{3}e^{-7D/2}\triangle e^{-D/2}
\eea
is the energy component of the energy-momentum tensor and $f^{\bot}_{(b)}$ is given by Eq.~(\ref{c-8}).
This energy density constraint determines both the diffeo-invariant time interval (\ref{1-14t}) and
the diffeo-invariant lapse function (\ref{ds-2n})
\be \label{1-11}
{\cal N}=\dfrac{\langle \sqrt{{\cal H}}\rangle}{\sqrt{{\cal H}}} 
\to \langle {\cal N}^{-1}\rangle \equiv 1.
\ee
The average of this energy density constraint~(\ref{1-7ab}) over the space volume
$\left\langle\dfrac{\delta W_{\rm H}}{\delta \ln N}\right\rangle=0$
leads to the energy constraint in the form of the Friedmann-type equation
\be \label{1-12}
\left[\dfrac{d\langle D\rangle}{d\tau}\right]^2=\langle \sqrt{{\cal H}}\rangle^2.
\ee
The solution of this equation yields the relation between the diffeo-invariant
time interval and the dilaton zeroth mode:
\be\label{1-13}
\tau-\tau_I=\int\limits_{\langle D\rangle_I}^{\langle D\rangle_0}d\langle D\rangle\langle
\sqrt{{\cal H}}\rangle^{-1}.
\ee
This means that the dilaton zeroth mode is a time-like variable.

In the Hamiltonian approach the energy constraint equation 
takes the form
\be \label{1-3n-1H}
P^2_{\langle D\rangle}-{\bf E}^2_{\rm U}=0,
\ee
where
\bea \label{1-3n-3H}
{\bf E}_{\rm U}=2\int\limits_{V_0}^{} \omega_{(1)}\wedge\omega_{(2)}\wedge\omega_{(3)}\sqrt{{\cal H}}.
\eea
We can treat quantity (\ref{1-3n-3H}) as the energy of the Universe in the field space of events.

The dilaton non-zeroth modes are not independent degrees of freedom, therefore
its velocity $v_{\overline{D}}$ and momentum $P_{\overline{D}}=2 v_{\overline{D}}$ are
equal zero
\be \label{min-3}
P_{\overline{D}}=0.
\ee
This equation is in agreement with the Dirac constraint of the minimality of three dimensional
hyper-surface at embedding it into the four dimensional Riemannian space~\cite{Dirac-1958}.

Thus, the constraint-shell Hilbert action takes the form
\bea\nonumber
&& W_{\rm c-shell}=\int d^4x p^j_{(a)}\partial_0{\bf e}_{(a)j}
+ \int dx^0 P_{\langle D\rangle}\partial_0\langle D\rangle]
\\ \nonumber
&& = \int\limits^{\langle D\rangle_0}_{\langle D\rangle_I} d \langle D\rangle
\\\label{csha-1} &&
\times \left[\!\int\limits_{V_0}^{}\! \omega_{(1)}\!\wedge\omega_{(2)}\!\wedge\omega_{(3)}
p^R_{(a)(b)}\omega^R_{(a)(b)} (\partial_{\langle D\rangle})\!\mp\! {\bf E}_U\!\right]\!\!,
\eea
where $\partial_{\langle D\rangle}=\dfrac{d}{d\langle D\rangle}=\langle \sqrt{{\cal H}}\rangle\dfrac{d}{d\tau}$.
The constraint-shell action requires D-diffeo-invariant initial finite volume and initial
data including an initial dilaton value $\langle D\rangle_I$ as certain input parameters given
at the instance of the Universe creation.
Here we collect and reproduce the known concepts just since we start from the standard Hilbert action.
The new facts are only the definitions of D-diffeo-invariant coordinates and variables with their
initial data. It is just the first goal of our approach to the Einstein GR. The next step is to compare
the classical choice of unstable dilaton value $\langle D\rangle_I=0$ and the infinite volume value
with their quantum values restricted by the uncertainty principle. We call this quantum theory
the quantum \textit{gravidynamics} (QGD) by analogy with electrodynamics and chromodynamics.
We have to define the region of validity of QGD and its predictable possibilities.

\subsection{The dilaton constraint}

The dilaton equation
\be\label{D-1}
\frac{\delta W_{\rm H}}{\delta D}=0
\ee
takes the form
\bea\label{e2t}
(\partial_\tau-{\cal N}_{(b)}\partial_{(b)})p_{{D}} = T_D,
\eea
where
\bea\label{e2dd}
T_D=\frac{4}{3}
e^{-D/2}\triangle[{\cal N}e^{-7D/2}]
- {\cal N}\partial_D{\cal H}. 
\eea
If the dilaton field is split into the zeroth mode and the non-zeroth ones 
$D=\langle D\rangle+\overline{D}$,
then equation~(\ref{e2t}) also is split into the equations for zeroth mode and the non-zeroth modes
restricted by constraint~(\ref{min-3}).

The solution of the zeroth dilaton mode equation
$\dfrac{\delta W_{\rm H}}{\delta \langle D\rangle}=0$ coincides with
the Friedmann-like solution of the energy constraint, see Eq.~(\ref{1-13}).

The equation for the non-zeroth modes 
\be\label{nD-1}
\frac{\delta W_{\rm H}}{\delta \overline{D}}=0
\ee
takes the form
\be \label{cm-DD}
T_{\overline{D}}=T_D-\langle T_{D}\rangle=0.
\ee
It determines $\overline{D}$ as one of static potentials.

Here we just adapted the Dirac--ADM Hamiltonian formulation~\cite{ADM-1959}
to the Maurer--Cartan forms. In terms of these forms, the Hilbert action
becomes a bilinear functional.
This means that the action of such a theory describes a physical system of the type of
a squeezed oscillator~\cite{Arbuzov:2010fz}.
It gives a hope to construct a quantum theory of such a system at the level of Cartan's forms
and to describe quantum processes. We treat the Einstein's General Relativity in terms of
diffeo-invariant variables and potentials as a candidate for formulation of
the Quantum Gravidynamics by the analogy with to Quantum Electrodynamics (QED).

%


\section{Quantum Gravidynamics}

\subsection{Formulation}

Quantum Electrodynamics is treated as the ideal to formulate any physical theory, including QGD.
Einstein used the Maxwell equations of classical electrodynamics as the example
of a same type field equations in the gravitation theory. QGD keeps this analogy.
If one neglects dynamical variables, in both the theories, there arise
the static Colombian type interactions of external sources.
In our case, if we neglect dynamical variables
(the luminosity $\langle D\rangle$ and the graviton $\omega^R$)
we obtain the black-hole type solution in the isotropic coordinates~\cite{Barbashov:2005hu}.
In the opposite case, far from external sources, the set of constraints and equations of the QED
describe only two transverse photons as independent degrees of freedom.
In our case far from heavy bodies the set of constraints and equations of QGD
describes only two  gravitons $\omega^R$ and the luminosity $\langle D\rangle=-\ln a=\ln(1+z) $,
where $z$ is identified with the redshift in the observational Cosmology.

The Einstein's cosmological principle~\cite{Einstein:1917ce} determines two classes of functions
distinguished by two projection operators. The first class determines the concept of the energy
of the Universe, while the second class determines the physical content of this energy
in the form of primordial gravitation waves and their vacuum expectation value.
If we neglect gravitons, the set of Einstein equations has only the trivial solution
$\langle D\rangle=0$. If we neglect the dilaton, the set of Einstein equations has also
the trivial solution $\omega^R=0$. However, both these approximations are only classical ones.
In quantum theory both trivial solutions are impossible due to the uncertainty principle.
To obtain consequences of the uncertainty principle in quantum gravidynamics we need
its perturbation theory generating functional.


\subsection{Generating functional of the quantum gravidynamics perturbation theory
\label{sect_4}}

In the Hamiltonian approach the energy constraint equation~(\ref{1-3n-1H})
can be quantized.
In quantum theory the canonical variables $\hat P_{\langle D\rangle}$ and ${\langle D\rangle}$
become operators with the commutation relation
$[\hat P_{\langle D\rangle}, {\langle D\rangle}]=\imath$,
and the energy constraint becomes the equation of the Wheeler--DeWitt type~\cite{DeWitt,Wheeler}
\be \label{5-C-7cc}
\left[\hat P^2_{\langle D\rangle}-{\bf E}^2_{\rm U}\right]
{\hat \Psi}_{\langle D\rangle_I, \langle D\rangle_0}=0.
\ee
By analogy with the unitary irreducible representation of the Poincar\'e group
in quantum field theory, we get a general operator solution of the Wheeler--DeWitt equation
for the universe as a sum of two exponents ordered with respect to the field evolution
parameter $\langle D\rangle$
\bea \label{09-07} \nonumber
&& {\hat \Psi}_{\langle D\rangle_I, \langle D\rangle_0} = {\hat A}_{\langle D\rangle_I}^+
{\hat{\mathbb{U}}}^{0}_{I}\frac{1}{{\sqrt{2{\bf E}_{0\rm U}}}}
+ {\hat A}_{\langle D\rangle_I}^- \frac{1}{{\sqrt{2{\bf E}_{0\rm U}}}}{\hat{\mathbb{U}}}^{I\dagger}_{0},
\\ \label{08-02}
&&{\hat{\mathbb{U}}}^{0}_{I}= T_{\langle D\rangle}
\exp \left\{-\imath\int\limits_{\langle D\rangle_I}^{\langle D\rangle_0}
d\langle D\rangle {\bf E}_{\rm U} \right\}.
\eea
It is a unitary operator of the universe evolution in the space of events
$[\langle D\rangle|F]$ relatively to the field evolution parameter
$[\langle D\rangle]$. The WDW functional
describes the creation of the universe at the time $\langle D\rangle_I$ and
its evolution from $\langle D\rangle_I$ till the present day moment $\langle D\rangle_0$.
The two terms correspond to the positive and negative energy, where
${\hat A}_{\langle D\rangle_I}^+$
can be interpreted as the operator of creation of the universe at the moment
$\langle D\rangle_I$,
and ${\hat A}_{\langle D\rangle_I}^-$ is the correspondent annihilation operator
with a commutation relation
$$[{\hat A}_{\langle D\rangle_I}^-,{\hat A}_{\langle D\rangle_I}^+]=1.$$
Negative energy is removed by the second quantization.
After the procedure of normal ordering the field Hamiltonian takes the form
\bea \label{08-06}
\widetilde{{\cal H}}&=&\rho_{\rm Cas}+:\widetilde{{\cal H}}:,
\eea
where
\bea \label{08-06N}
\rho_{\rm Cas}=\sum\limits_{{\bf k}\neq 0}^{} \dfrac{\omega_{{\bf k}}}{2}f_{{\bf k}}
\eea
is the density of the Casimir energy of gravitons with the one particle energy $\omega_{{\bf k}}$
that can be determined in finite space by the analogy with the Casimir energy of
photons~\cite{Casimir-48,Bordagempty-6}.
The divergent integral in the Casimir energy can be regularized by introduction
of a simple distribution function
\bea
&& f_{\bf k}=\left[1- \exp\left(d\sqrt{{\bf k}^2}\right)\right]^{-1},
\eea
where $d$ is an input parameter as the volume dimension.
This distribution function reflects the degeneration of the vacuum state
over momenta~\footnote{The set of the vacuum degeneration parameters can be treated
as a von Neumann's ``Statistishe Gesamtheinten''~\cite{vonN-1932}
or the Blokhintsev's ``Quantum Ensemble''~\cite{bdi-1940}.}.

Recall that the Casimir energy is the result of normal ordering of the product 
of field operators in the free Hamiltonian. This means that the graviton occupation 
number cannot be equal zero.
It is natural to suppose that at the Beginning the Casimir energy of the Universe
\bea \label{08-07}
{\bf E}^I_{\rm U}=2\int\limits_{V_0}^{} d^3x \sqrt{\rho_{\rm Cas}}
\eea
dominated. Following Kasner~\cite{kasner-1921}, we suppose that the region of validity
of the GR is the creation and evolution of the Universe as the whole.

Moreover, the key assumption of the our model is the equality of the Casimir dimension $d(a)$
and  the Universe size defined as the horizon:
\be \label{d-r}
d(a)=r(a), \qquad a=e^{\langle D\rangle}.
\ee
The Casimir energy of the Universe in the volume $V_0=\dfrac{4\pi}{3H_0^3}$
is proportional to the Hubble constant $H_0$. The solution of equation~(\ref{d-r})
given in refs.~\cite{Pervushin:2011gz,Arbuzov:2010fz} corresponds to the rigid state
$p=+1\cdot \rho$. The rigid state dominance is in agreement with the SNe~Ia
data~\cite{Perlmutter,Riess}, in the framework of conformal GR where the long conformal
intervals are identified with observational distances~\cite{Behnke:2001nw,Zakharov:2010nf}.
This agreement means that the long-distance SNe~Ia data testify to an almost empty Universe
during all time of its evolution with the dominant energy density contribution from the
Casimir energy.

The rigid state dominance means that SNe~Ia data correspond to the primordial value 
of the Hubble parameter and the dilaton initial data $\langle D\rangle_I$ at the Planck scale,
where the Hubble parameter value coincides with the Planck mass one.
In this case the primordial value of the Hubble parameter as the input parameter and
the set of Eqs.~(\ref{cm-D})---(\ref{cm-DD}) completely defines the perturbation theory.

\subsection{Perturbation theory}

An expansion of the evolution generator can be constructed as
\bea \label{08-08}
{\bf E}_{\rm U} &=&  2\int d^3x \sqrt{ \rho_{\rm Cas} + :\widetilde{{\cal H}}:}
\nonumber \\
&=& {\bf E}^I_{\rm U}+\frac{{\bf H}_{\rm QFT}}{\sqrt{\rho_{\rm Cas}}}+\cdots
\\ \nonumber
{\bf E}^I_{\rm U} &=& 2\int d^3x \sqrt{\rho_{\rm Cas}},
\\ \nonumber
{\bf H}_{\rm QFT} &=& \int d^3x:\widetilde{{\cal H}}.
\eea
We can see that this expansion coincides with the quantum field perturbation theory
\bea  \nonumber \label{08-10}
\hat{\mathbb{U}} = \underbrace{{\mathbb{U}}_{\rm Cas}}_{\rm Casimir~ miniuniverse}
\cdot \underbrace{T_{\widetilde{t}}\exp \left\{-\imath \int\limits_{\widetilde{t}_I}^{\widetilde{t}_0}
d\widetilde{t} {\bf H}_{\rm QFT}\right\}}_{ S-matrix},
\eea
where $d\widetilde{t}=\dfrac{d\langle D\rangle}{\sqrt{\rho_{\rm Cas}}}$
coincides with the conformal time interval as it will be shown later.
This expansion is made under the assumption of the dominance of the
Casimir energy of graviton.

\section{Vacuum Energy of Diffeo-Invariant Gravitons}

\subsection{Quantum graviton}

One can see that the Hilbert action contains the global dilaton part~(\ref{1-3a})
(that solves the time-energy problem in the complete theory), the
graviton part~(\ref{1-4a}) (that describes diffeo-invariant transverse gravitons),
and the potential part~(\ref{1-4ab}) (that describes the black hole type solution
induced by heavy massive bodies).

If there are no massive bodies at the Beginning of Universe, then the last
term~(\ref{1-4ab}) can be neglected. In this case the second terms describes
an exactly solvable model of diffeo-invariant transverse gravitons
of the type of the Bondy--Trautmann ones~\cite{Pervushin:2011gz}:
\bea \label{R-1a}
&& \omega_{(a)(b)}^R(\partial_{(c)})=\frac{\imath\sqrt{3}}{V_0}\sum_{\textbf{k}^2\not=0}
\dfrac{e^{\imath\textbf{\textbf{k}\textbf{X}}}}{\sqrt{2\omega_{\textbf{k}}}}
\textbf{k}_{c}g^R_{\textbf{k}(a)(b)},
\\ \nonumber
&& g^R_{\textbf{k}(a)(b)}= 
[\varepsilon^{R(A)}_{(a)(b)}(\textbf{k})
g_{\textbf{k}}^{(A)+}(\tau)+\varepsilon^{R(A)}_{(a)(b)}(-\textbf{k})
g_{-\textbf{k}}^{(A)-}(\tau)].
\eea
Here we introduced the polarization tensors
\bea \nonumber
&& {\varepsilon}^{R(1)}_{(a)\,(b)}(\textbf{k})
= \frac{1}{\sqrt{2}}[\varepsilon_{(a)1}(\textbf{k})\varepsilon_{(b)2}(\textbf{k})
+ \varepsilon_{(a)2}(\textbf{k})\varepsilon_{(b)1}(\textbf{k})],
\\ \nonumber
&&{\varepsilon}^{R(2)}_{(a)\,(b)}(\textbf{k})
= \frac{1}{\sqrt{2}}[\varepsilon_{(a)1}(\textbf{k})\varepsilon_{(b)1}(\textbf{k})
- \varepsilon_{(a)2}(\textbf{k})\varepsilon_{(b)2}(\textbf{k})].
\eea
The Hilbert action~(\ref{1-2})
is a bilinear functional of diffeo-invariant gravitons. Therefore,
in the region far from heavy bodies we can consider the gravitons
on an equal footing with the Standard Model quantum fields.

The graviton Hamiltonian takes the form
\bea \label{2-10gh}
\textsf{H}^g_{\tau}=\sum\limits_{\textbf{k}^2\not =0}
\dfrac{p_{\textbf{k}}^gp_{-\textbf{k}}^g+ e^{-4\langle D\rangle}
\textbf{k}^2 \overline{g}_{\textbf{k}}\overline{g}_{-\textbf{k}}}{2}.
\eea
Solutions of the corresponding graviton equations can be obtained
in terms of the conformal time interval~\cite{Pervushin:2011gz}
$d\eta =a^2(\tau)d\tau$, where $a=e^{-\langle D\rangle}$ and
\be \label{2-15gh}
g_{\textbf{k}}^{A\pm}
=g^{A\pm}_{I\textbf{k}}\cdot \exp \{\pm\imath |{\bf k}|(\eta-\eta_I)  \}
\ee
These solutions describe standard free quantum fields in a finite volume.
The Hamiltonian vacuum expectation value yields the Casimir energy with 
the single input parameter $H_0$ given by the SNe~Ia data~\cite{Perlmutter,Riess}.

\subsection{The single graviton case}


We consider the case of the metric
\bea \label{H-2a-t1}
{ds}^2 = d\eta^2 -dX_{(3)}- [{\omega}^{2}_{(1)} + {\omega}_{(2)}^2] 
\eea
when the spin-connection coefficients take the form 
\bea \label{H-2a-s1}
\omega_{(1)(1)}(d)&=&-\omega_{(2)(2)}(d)=d D_{1}(\eta_{(-)}),\\\label{H-2a-ss1}
\omega_{(1)(2)}(d)&=&\omega_{(2)(1)}(d)= d D_{2}(\eta_{(-)}).
\eea
Here the functions $D_{1}, D_{2}$ depend on only the light cone coordinate
\be \label{H-2a-lc1}
\eta_{(-)}=\eta-dX_{(3)}.
\ee
It is just the case of a strong gravitation wave, 
when the spin-connection coefficients are the solution of 
the exact Einstein equations for the metric~(\ref{H-2a-t1}).

Let us pass to the complex coordinates, linear differential forms and fields
\bea\label{C-t1}
z&=&x^1+ix^2,\qquad Z(D)=X_{(1)}+i X_{(2)}, 
\nonumber \\
D&=&D_1+iD_2,\qquad \omega(d)={\omega}_{(1)}+i{\omega}_{(2)}.
\eea
This means that the integral form $Z^*(D$ of the tangent coordinates can be obtained 
via the Eq.~(\ref{1-16aabb})
\bea \label{C-t5}
\frac{dZ(D)}{d D}=Z^*(D), \qquad Z(0)=Z^0
\eea
that allows us to calculate their time dependence in the field of the strong
gravitation wave $D(\eta_{(-)})$. The solution of Eq.~(\ref{C-t5})
\be\label{C-t6}
Z(D)=R(D) e^{i\theta(D)}
\ee
describes also the rotations of any vectors in the complex plane~(\ref{C-t1}),
where all measurable quantities are diffeo-invariants, including the photon spin (polarization).
The substitution of (\ref{C-t6}) into in Eq.~(\ref{C-t5}) yields
\bea\nonumber
&&e^{i2\theta(D)} d \left[\ln \frac{R(D)}{R(0)}+i\theta(D)\right]
\\ \nonumber
&=& \cos2\theta d\ln\frac{R}{R_0} +\frac{1}{2}d \cos2\theta 
\\ \nonumber
&+& i \sin2\theta d\ln\frac{R}{R_0} +i \frac{1}{2}d \sin2\theta
\\ \label{C-t7}
&=& d \left[D_1+iD_2\right].
\eea
Using the independence condition $dX_{(1)}/dX_{(2)}=0$ one can write the solution 
of these equations in the form
\bea\label{C-t8}
R(D)&=&R_0\exp\left\{\sqrt{\widetilde{D}^2_1+\widetilde{D}^2_2} 
-\frac{1}{2}\right\},
\\ \label{C-t9}
\cos2\theta(D)&=&\frac{\widetilde{D}_1}{\sqrt{\widetilde{D}^2_1+\widetilde{D}^2_2}},
\\ \label{C-t10}
\sin2\theta(D)&=&\frac{\widetilde{D}_2}{\sqrt{\widetilde{D}^2_1+\widetilde{D}^2_2}},
\\ \nonumber
\widetilde{D}_1&=&{D}_1+\frac{1}{2}\cos2\theta_0,
\\ \nonumber
\widetilde{D}_2&=&{D}_1+\frac{1}{2}\sin2\theta_0.
\eea
Here $R_0,\theta_0$ are initial data compatible with the initial values
\bea\label{C-t11}
R(D=0)&=&{R}_0,
\\ \label{C-t12}
\cos2\theta(D=0)&=&\cos2\theta_0,
\\ \label{C-t13}
\sin2\theta(D=0)&=&\sin2\theta_0.
\eea
and the perturbation series
\bea\nonumber
Z(0) = Z_0, \qquad Z(D) = Z_0+Z^*_0D + O(D^2).
\eea

%

\subsection{The CMB temperature test of the space-time~foam in QGD}

The Einstein interval in the Friedmann homogeneous approximation
\be \label{f-4}
\left[\frac{ds}{dt}\right]^2=1-a^2(t)\left(\dfrac{d{\bf X}}{dt}\right)^2=[1-{\bf v}^2_\eta]
\ee
defines the cosmic evolution of photon velocities
$$\dfrac{d{\bf X}}{dt}=\dfrac{d{\bf X}}{ad\eta}=\dfrac{{\bf v}_\eta}{a}$$
and, therefore, the cosmic evolution of the CMB photon temperature
\be\label{tf-4}
T^{\rm F}_{\rm CMB}= \frac{T_{0\rm CMB}}{a(t)},\qquad T_{0\rm CMB}=2.725 {\rm K}.
\ee
The QGD basic approximation identifies observational velocities with
the conformal interval ones obtained as the squired interval vacuum expectation value
\bea \label{cf-4}
&& \left\langle 0 \left|\dfrac{\widetilde{ds}^2}{(d\eta)^2}\right|0 \right
\rangle_{\rm space-time~foam}
\\\nonumber &&
= 1 - \underbrace{|{\bf v}|^2}_{T_{0\rm CMB}} - \underbrace{X_{(a)}X_{(b)}
\langle 0| \omega^R_{(a)(b)}(\partial_\tau)\omega^R_{(a)(b)}(\partial_\tau) |0\rangle}_{
{3|\bf X}|^2 H_0^2 \cdot \Omega_{\rm gr}(1+z)^2=\triangle T_{\rm CMB}(z)},
\eea
where $\Omega_{\rm gr}$ is of order of unit.

Instead of the Friedmann CMB temperature~(\ref{tf-4}),
the QGD basic approximation yields the CMB temperature evolution
\be \label{tcf-4}
T^{\rm QGD}_{\rm CMB}= {T_{0\rm CMB}} + \Delta{T_{0\rm CMB}}(1+z)^2.
\ee
If the present day ratio ${\Delta T_{0\rm CMB}}/{T_{0\rm CMB}}\sim 10^{-6}$,
than the QGD basic approximation predicts that at the time of the recombination
$z_{\rm recomb}\sim 10^3$,
when the CMB radiation decouples from the massive matter,
the quantum space-time foam anisotropy 
was of order of the isotropic part ${T_{0\rm CMB}}$.
This means that at the time of the recombination the quantum graviton space-time foam anisotropy
can be able to form galaxies and large-scale massive matter structures in the Universe.
At the present day time $z_0\ll 1$ the QGD basic approximation yields the flat conformal interval
that is identified with the measurable distance
in the Dirac version~\cite{Dirac:1973gk} of the Einstein theory.
In this conformal version the temperature history of the Universe becomes the history of
elementary particle masses. The origin of these masses and their cosmic evolution are
the objects of the next Sections.

\section{Observational Arguments
\label{sect_7}}

\subsection{SNe Ia data as an evidence of vacuum energy dominance
\label{sect_7.2}}

Observations of the so-called standard candles supernovae~\cite{Perlmutter,Riess}
demonstrated that they are turned out to be too far away from us. Cosmologists had
to introduce one more inflation in the Universe evolution scenario,
explaining this inflation by the $\Lambda$ term or by a modification of the GR.
Meanwhile, we have shown~\cite{Behnke:2001nw,Zakharov:2010nf}
that the observed supernova data (SNe Ia) can be explained if one just switches to conformal
variables in the standard Friedmann equations. Fitting the data in this way shows that the
dominant contribution to the Universe energy density should have the rigid equation of state
$p=+1\cdot \rho$. In general, the rigid state can be realized in different ways, {\it e.g.} with
the help of an additional scalar field, see ref.~\cite{Zakharov:2010nf}.
Here we adapt the possibility to identify the dominant rigid state with the Casimir 
vacuum energy as it was suggested in ref.~\cite{Pervushin:2011gz}.
In fact, the vacuum energy dominance corresponds to the rigid state with the cosmological 
evolution
\bea\label{h-1}
H(z)  &=& \frac{d \langle D\rangle}{d\eta}=\frac{H_0}{(1+z)^2}, \\ \label{ha-1}
(1+z) &=& e^{\langle D\rangle},
\\ \label{ha-2}
d\eta &=& \frac{d \tau}{(1+z)^2},
\eea
where $d\eta$ is the conformal time and $H_0$ is the present day value of the Hubble constant.

\subsection{The Planck scale cosmic hierarchy
\label{sect_7.1}}

The Planck epoch corresponds to the redshift value $z_I$ for which the primordial
Planck mass value
\be\label{dl-3}
M^*_{I\rm Pl}\equiv M^*_{\rm Pl}(z_I)= M^*_{0\rm Pl}(1+z_I)^2
\ee
coincides with the the primordial Hubble constant
\be\label{dh-3}
H_I\equiv H(z_I)=M^*_{I\rm Pl}= M^*_{0\rm Pl}(1+z_I)^2. 
\ee
Thus the value of the primordial redshift is of the order
\be \label{z-1}
z_I \simeq 10^{15}.
\ee
As discussed in~\cite{Pervushin:2011gz},
this definition of the Planck epoch follows also from the Planck least action
principle applied at the initial moment of the universe evolution.

So, it is quite natural to assume that at the initial moment of a quantum
universe creation there is just a single energy scale introduced as the initial
data. The present Hubble constant value and the Planck mass are related to each
other just by the age of the Universe expressed in terms of the redshift.
The evolution of an energy scale of a physical quantity from the initial moment
to the present day is obviously defined by the corresponding dimension $d$
and conformal weight $w$. In particular besides the evolution of the Hubble constant
and the Planck mass, we can consider evolution of a particle mass which takes today
the value $m_0$
\bea\label{h-set-1}
H_{0} &\simeq& H_I\cdot {(1+z_{\rm I})}^{-2}, \quad \{d=1,w=2\},
\\ \label{r-set-1}
m_0 &\simeq& H_I \cdot {(1+z_{\rm I})}^{1}, \quad \{d=1,w=-1\},
\\ \label{p-set-1}
M^*_{\rm Pl~0} &\simeq& H_I\cdot {(1+z_{\rm I})^{2}},\quad \{d=1,w=-2\}.
\eea
Using the known values of $H_0$ and $M_{\mathrm{Planck}}$ together with the defined
above magnitude of $z_i$, we see that $m_0\simeq 300$~GeV, {\it i.e.} it is just of
the order of the electroweak energy scale.
Obviously, we still have to demonstrate how does a massive particle emerge in the conformally
symmetric model and how its mass is related to the {\em primary} energy scale $H_I$.
It is worth to note that in our construction $m_0$ provides the order of the
maximal possible mass of an elementary particle in the given universe,
in accordance with the hypothesis about the so-called {\em maximon}
proposed by M.A.~Markov~\cite{Markov:1965max,Markov:1994hi}.

We limit ourselves by the set of particles existing in the Standard Model.
Obviously, the electroweak scale is related both to the Higgs boson mass
(and its vacuum expectation value)
and to the top quark mass. Note that the coincidence of the these energy scales is
an unresolved puzzle within the SM.
Here we assume that the mechanism of dimensional transmutation~\cite{Coleman:1973jx}
is working in the sector of the SM, which contains the most intensive interactions
of this model, see the next Section.
At the Planck epoch renormalization energy scale is naturally given by $H_I$.
The Coleman-Weinberg mechanism then provides non-zero masses and condensates for
both scalar and fermion fields. Since the two coupling constants are both of the
order of unity, the emerged masses and condensates are of the same order as $H_I$.
In particular using Eq.~(\ref{h-set-1}), we can relate the top quark mass with
the Hubble constant and the Planck mass:
\be
m_t \simeq H_0 [1+z_I]^3, \qquad m_t \simeq M^*_{\rm Pl~0} [1+z_I]^{-1}.
\ee
On the other hand, the top quark condensate should have the same
energy scale\footnote{Fermion condensates are negative {\em by construction} since
they are integrals over a closed fermion loop.}, so
\be \label{1-h}
\langle \bar t t\rangle= - \gamma_t m^3_{t}
\ee
where $\gamma_t$ is a dimensionless constant of the order of unity.
To get a concrete number for parameter $\gamma_t$, we have to regularize the divergent
integral. It can be done by introduction of a simple distribution function:
\bea
&& \gamma_t = - \frac{\langle \bar t\, t\rangle}{m^3_{t}}
= 4N_c\int \frac{d^3{\bf p}}{(2\pi)^3}\frac{1}{2\sqrt{{\bf p}^2+1}}f_{1,+}({\bf p}) \approx 0.39,
\nonumber \\
&& f_{1,\pm}({\bf p})=\left[1\pm \exp\left(\sqrt{{\bf p}^2+1}-1\right)\right]^{-1}.
\eea
This distribution function reflects the degeneration of the vacuum state
over momenta. The set of the vacuum degeneration parameters can be treated
as von Neumann's ``Statistishe Gesamtheinten'' \cite{vonN-1932}
or the Blokhintsev's ``Quantum Ensemble'' \cite{bdi-1940}.

\subsection{Conformal symmetry breaking in the Standard Model
\label{sect_4.4}}

In ref.~\cite{Pervushin:2012dt} a simple reduction of the SM to a conformal-invariant
theory was suggested. As shown in~\cite{Arbuzov:2014rra,Arbuzov:2016xte}, 
the infrared instability of the theory leads to quantum anomalies which 
break the conformal symmetry spontaneously.
This effect is clearly seen in the one-loop approximation for effective potential
of the Higgs boson. We briefly repeat the main steps of this construction 
and extend it by using a direct calculation of the top quark condensate value.

Let us look at the Higgs boson sector of the SM taken without the tachyon mass term.
For the first approximation we take the most intensive terms of Higgs boson interactions,
{\it i.e.} the self-interaction and the the Yukawa coupling with the top quark
\be\label{L_int1}
{\cal L}_{\mathrm{int}}(\phi) \approx - \frac{\lambda}{4}\phi^4 - \frac{y_t}{2} \phi~\bar{t}t.
\ee
Note that we have here already the single neutral component of the primary complex doublet
scalar field.
Following the standard Brout--Englert--Higgs mechanism 
(and the Einstein's cosmological principle),
we split $\phi$ into its mean field $\langle\phi\rangle$ and 
particle-like non-zeroth harmonics $h$
\bea\label{mf-4}
\phi= \langle\phi\rangle + h, \quad \int d^3x h=0.
\eea
In the same manner, normal ordering of the fermion pair in the Yukawa interaction term
in Eq.(\ref{L_int1}) can be decomposed as
$\bar t\, t = :\bar{t}\, t: + \langle \bar{t}\, t \rangle$.
By construction $m_t=(y_t/\sqrt{2}) \langle\phi\rangle$ where $y_t\approx 0.99$ is the Yukawa
coupling of the top quark, and $v\approx 246.22$~GeV is the Higgs boson
vacuum expectation value.

One can see that the top quark condensate density
supersedes the phenomenological negative square mass term in the Higgs potential.
As a consequence, we have a non trivial minimum in the Higgs field potential
\bea \nonumber
V(\phi) &=& \frac{\lambda}{4}\phi^4 + \frac{y_t}{\sqrt{2}} \phi \langle \bar t\, t \rangle,
\\ \nonumber
\left.\frac{dV(\phi)}{d\phi}\right|_{\phi=v}&=&\lambda v^3
+ \frac{y_t}{\sqrt{2}}\langle \bar t\, t\rangle = 0,
\\ \label{h_mass1}
\frac{d^2V(\phi)}{d\phi^2}|_{\phi=v}&=&m^2_{h } = 3\lambda v^2.
\eea
So the Higgs particle mass in the tree-level approximation is defined as
\bea \label{f-3}
m^2_{0,h}=-\frac{3y_t\langle \bar t\, t\rangle}{\sqrt{2}\, v}\approx 131\ {\mathrm{GeV}},
\eea
where we used Eq.~(\ref{1-h}) to define the top quark condensate value
$\langle \bar t\, t\rangle \approx - ( 126 \ {\mathrm{GeV}})^3$
via the known $m_t\approx 173$~GeV.
One can see that the standard Brout--Englert--Higgs mechanism is reproduced except
that the value of the self-coupling constant $\lambda$ is $3/2$ times less than
in the SM.
Note that this quantity will be measured experimentally only at high-precision
future linear $e^+e^-$ colliders. Note also the there is no any experimental
data on the values of heavy quark condensates since they do not contribute
to hadronic observables.

The above results are obtained in the lowest semi-classical
approximation. Certainly, they can be shifted by radiative corrections.
But, contrary to the case of the Standard Model, we do not have quadratically
divergent corrections to the Higgs boson mass because they are cured by
the conformal symmetry in the classical Lagrangian.

\section{Conclusions
\label{sect_8}}

We have shown that the Einstein theory supplemented by
the electroweak Standard Model in terms of diffeo-invariant coordinates
and variables predicts not only relativistic effects, such as the
Mercury anomaly or the double Eddington's angle of the photon deviation
in the Sun gravitational field,
but also the recent observational facts in Cosmology and the set of quantum
effects. They include:

 1. The arrow of time arises due to the vacuum postulate at the Planck scale
 $$(1+z_I)=\left[\dfrac{M^*_{\rm Pl}}{H_0}\right]^{{1}/{4}}\approx 10^{15}.$$

 2. The SNe Ia data justify the dominance of the Casimir vacuum energy with
the rigid-state cosmic evolution and the primordial Hubble parameter value
$H_I=H_0 (1+z_I)^2$.

 3. The evolution of the vacuum condensates of elementary fields yields
the proper order of the Higgs boson mass $M_{\rm Higgs}\sim H_0 (1+z_I)^3$. 

 4. The intensive vacuum creation of electroweak bosons~\cite{Pervushin:2011gz} 
and their consequent decays yield the present day number of the CMB photons
$N_\gamma \approx \alpha^2_W \cdot (1+z_I)^6\simeq 10^{87}$.

The presented above model for the definition of the Planck epoch
is based on the assumption that the conformal symmetry is a fundamental
property of Nature. A spontaneous breaking of this symmetry due to
a quantum conformal anomaly provides a single energy scale. We have
shown that the energy scales of the Hubble constant, the Planck mass,
and the SM can be received by the cosmological evolution from the
single scale taking into account different conformal weights of the
corresponding quantities.
As concerning the origin of the QCD energy scale, one can try
to look for its relation to the electroweak one, but that goes
beyond the scope of this article.

Some other aspects of the discussed above model were considered
in our earlier papers~\cite{Barbashov:2005hu,Arbuzov:2010fz,Pervushin:2011gz,GC-2010}.
Further work on construction and verification of the model is in order.

It is interesting to perform a super-analog of the Borisov-Ogievetsky
nonlinear realization~\cite{Borisov:1974bn} in order to obtain the
corresponding number of Goldstone fields for both the general relativity
and the minimal Standard Model of the electroweak and strong interactions
(10 tensors, 48 vectors, 96 fermions, and 4 scalars).
We have shown in ref.~\cite{Diego-2015} that this super-nonlinear realization as
the last unification of the GR and SM can be based on the quaternionic
super-extension of the supertwistor construction given in ref.~\cite{Litov-84}.
According to this quaternionic supertwistor construction, the number of fermionic
fields can be extended beyond the pure supertwistor one with coherent
super-quaternionic states spanning the super-Hilbert spaces.

\section*{Acknowledgment}

The authors are grateful for useful discussions to participants of
the seminars of the Institute for Gravitation and Cosmology,
the People's Friendship University of Russia (PFUR), Moscow, and
the Russian Gravitational Association (VNIIMS), Moscow.
A. Pavlov is grateful to the Joint Institute for Nuclear Research
for hospitality.
NSH was supported by Nafosted undes grant number 103.03-2012.02.
A. Arbuzov thanks the Dynasty foundation for a 

\end{document}